\documentclass[a4paper,12pt]{article}
\usepackage[dvips]{graphicx}
\DeclareGraphicsExtensions{.pdf,.jpg}

\usepackage{amsfonts,amsmath,amsthm,amssymb}
\usepackage{mathrsfs}
\usepackage[onehalfspacing]{setspace}
\author{L. Ort\'{i}z}

\title{Thermal state on a cylindrical spacetime}

\begin{document}

\maketitle

\begin{center}

Department of Mathematics\\
The University of York\\
York YO10 5DD, U. K.\\\vspace{0.4cm}

\normalsize{\textbf{Abstract}}\\\end{center} \small{We proof that
if we have a thermal equilibrium state on Minkowski spacetime in
two dimensions then we have a thermal equilibrium state on the
cylindrical spacetime obtained from this Minkowski spacetime by
making $2\pi$-periodic the spatial direction. We perform this by
using the algebraic approach to Quantum Field Theory.}

\section{Introduction}

\normalsize{Quantum} Field Theory on non simply connected spaces
has been studied before by several authors \cite{rBasDow79},
\cite{deWittHartIsha79}, \cite{crCrambsKay96}. However, as far as
we know, none of them have used the algebraic approach to quantum
field theory in the sense of Haag and Kastler \cite{rHaagdKast64}.
Here we shall address this problem by using the generalization of
Algebraic Quantum Field Theory (AQFT) \cite{rHaagdKast64} given by
Brunetti, Fredenhagen and Verch \cite{rBrukFrerVer03}. We take as
a non simply connected space a cylinder and a simply connected
space a plane.

We will say that a state $\omega$ is a thermal equilibrium state
at temperature $\textsf{T}$ if it satisfies the KMS condition
\begin{equation}\nonumber
\omega(B(\alpha_{t}A))=\omega((\alpha_{t-i\beta}A)B),
\end{equation}
where $A$ and $B$ are two elements of the algebra on which
$\omega$ is defined, $\alpha_{t}$ is the automorphism on the
algebra corresponding to translations in time and
$\beta=\frac{1}{\textsf{T}}$. In this work we shall proof that if
this condition is satisfied by a state in two dimensional
Minkowski spacetime then it is satisfied for a state defined on a
cylindrical spacetime obtained from Minkowski spacetime by making
the spatial direction $2\pi$-periodic. The relation between the
two states will be specified below.

The organization of this paper is as follows. In section 2, we
proof that if the KMS condition is satisfied for a state in
Minkowski spacetime then it is satisfied for the corresponding
state in the cylindrical spacetime. In section 3, we discuss this
result just by using the formalism introduced by Haag and Kastler
and we compare our result with the image method for analyzing the
same problem.

\section{Thermal state on a cylinder and on a plane}

A natural mathematical concept we can use for our purposes is the
concept of covering space. Let us spell out how this concept
enters in our problem. If we consider
$\mathbb{R}^{1}\times\mathbb{R}^{1}$ as the covering space of
$\mathbb{R}^{1}\times\mathbb{S}^{1}$ then $\forall$
$x\in\mathbb{R}^{1}\times\mathbb{S}^{1}$ there is a neighborhood
$V$ of $x$ such that $\pi^{-1}\left( V \right)$ is a family $\{
U_{\alpha}\}$ of open disjoint pairwise subsets of
$\mathbb{R}^{1}\times\mathbb{R}^{1}$ and
$\pi:U_{\alpha}\rightarrow V$ is a homeomorphism of $U_{\alpha}$
to $V$.
$\pi:\mathbb{R}^{1}\times\mathbb{R}^{1}\rightarrow\mathbb{R}^{1}\times\mathbb{S}^{1}$
is called the covering map.

In \cite{rBrukFrerVer03}, which we will refer to as BFV, the
starting point is to consider the category of all globally
hyperbolic spacetimes, $\mathfrak{Man}$, with morphisms, $\psi$,
the isometric embeddings between two of these spacetimes, the
objects of the category. An isometric embedding is a map
$\psi:\mathcal{M}_{1}\rightarrow\mathcal{M}_{2}$, where
$\mathcal{M}_{1}$ and $\mathcal{M}_{2}$ are globally hyperbolic
spacetimes such that $\psi$ is a diffeomorphism onto its range and
$\psi$ is an isometry, $\psi_{*}\mathbf{g}_{1}=\mathbf{g}_{2}$
when $\psi$ is restricted to $\mathcal{M}_{1}$.

We can apply this concept to our problem as follows: Obviously
$\mathbb{R}^{1}\times\mathbb{S}^{1}$ and
$\mathbb{R}^{1}\times\mathbb{R}^{1}$ are not diffeomorphic but if
we just consider a small diamond shaped region, $\mathcal{D}_{c}$,
on $\mathbb{R}^{1}\times\mathbb{S}^{1}$ then under the covering
map this region maps to an infinite denumerable family
$\{\mathcal{D}_{i}\}$, $i=0,\pm 1, \pm 2 ...$ of diamond shaped
regions in $\mathbb{R}^{1}\times\mathbb{R}^{1}$. Clearly in each
element, say $i=0$, of this family the covering map induces an
isometric embedding which pushforward the metric on
$\mathcal{D}_{c}$ to $\mathcal{D}_{p}$ where $\mathcal{D}_{p}$ is
the diamond shaped region in $\mathbb{R}^{1}\times\mathbb{R}^{1}$
which corresponds to $i=0$. The way that the covering map induces
an isometric embedding from $\mathcal{D}_{c}$ to $\mathcal{D}_{p}$
can be seen more clearly if we introduce atlases
$\{(U_{\alpha},u_{\alpha})\}$ and $\{(V_{\beta},v_{\beta})\}$ in
these two manifolds. Then $\pi^{-1}$ determines continuous maps
\cite{wGresHalrVan72}
\begin{equation}\label{E:136a}
\pi^{-1}_{\beta\alpha}:u_{\alpha}(U_{\alpha}\cap\pi(V_{\beta}))\rightarrow
v_{\beta}(V_{\beta})
\end{equation}
where $\pi^{-1}_{\beta\alpha}=v_{\beta}\circ\pi^{-1}\circ
u_{\alpha}^{-1}$. In the present case we can cover
$\mathcal{D}_{c}$ and $\mathcal{D}_{p}$ with the same single
chart. Hence in this case the maps (\ref{E:136a}) are smooth and
$\pi^{-1}$ is smooth. So we have a diffeomorphism between
$\mathcal{D}_{c}$ and $\mathcal{D}_{p}$. Clearly we can make it an
isometric embedding by pushing forward the flat metric on the
cylinder to the flat metric on the plane. This does not depend on
the element we take in the family $\{\mathcal{D}_{i}\}$, $i=0,\pm
1, \pm 2 ...$. If we take other element in $\{\mathcal{D}_{i}\}$
then we can relate it to $\mathcal{D}_{p}$ by acting on
$\mathcal{D}_{p}$ with an element $\gamma_{n}$ of $\Gamma$ where
$\Gamma$ is the discrete abelian group of spatial translations by
$2\pi$ in $\mathbb{R}^{1}\times\mathbb{R}^{1}$. In these
circumstances the following diagram commute
\begin{equation}
\begin{array}{ccc}
\mathcal{D}_{c} & \xrightarrow{\pi^{-1}} & \mathcal{D}_{p}  \\
\pi^{-1}\searrow &  & \swarrow\gamma_{n}  \\
 & \mathcal{D}_{n} &
\end{array}
\end{equation}
where $\mathcal{D}_{n}:=\gamma_{n}\mathcal{D}_{p}$.

At this stage we can apply the formalism given by BFV with the
diamond shaped regions introduced above as the elements of
$\mathfrak{Man}$. Let us write down explicitly the elements which
are relevant for our purposes.

In addition to the category $\mathfrak{Man}$ above introduced we
need to introduce the category $\mathfrak{Alg}$ whose objects are
all the C*-algebras, and the morphisms, $\alpha$, are faithful
unit-preserving *-homomorphisms. Then a locally covariant quantum
field theory is a covariant functor $\mathscr{A}$ between the
categories $\mathfrak{Man}$ and $\mathfrak{Alg}$, in a diagram we
have
\begin{equation}
\begin{array}{ccc}
(M,\mathbf{g}) & \xrightarrow{\psi} & (M',\mathbf{g}')  \\
\mathscr{A}\downarrow &  & \downarrow\mathscr{A}  \\
\mathscr{A}(M,\mathbf{g}) & \xrightarrow{\alpha_{\psi}} &
\mathscr{A}(M',\mathbf{g}')
\end{array}
\end{equation}
together with the covariance properties
$\alpha_{\psi'}\circ\alpha_{\psi}=\alpha_{\psi'\circ\psi}$ and
$\alpha_{idM}=id_{\mathscr{A}(M,\mathbf{g})}$
for all morphism
$\psi\in\textrm{hom}_{\mathfrak{Man}}((M_{1},\mathbf{g}_{1})(M_{2},\mathbf{g}_{2}))$,
$\psi'\in\textrm{hom}_{\mathfrak{Man}}((M_{1},\mathbf{g}_{1})(M_{2},\mathbf{g}_{2}))$
and all $(M,\mathbf{g})\in\textrm{Obj}(\mathfrak{Man})$. There are
two additional properties which are satisfied by the functor
$\mathscr{A}$ \cite{rBrukFrerVer03}, but for our purposes it is
enough with the property just introduced. We should note that in
our problem $M$ corresponds to the region $\mathcal{D}_{c}$ or
$\mathcal{D}_{i}$.

Also we need to introduce one category more, the category of the
set of states which we will denote as $\mathfrak{Sts}$. An object
$\mathbf{S}$ $\in$ $\textrm{Obj}(\mathfrak{Sts})$ is a set of
states on a C*-algebra $\mathcal{A}$. Morphisms between members
$\mathbf{S}'$ and $\mathbf{S}$ of $\textrm{Obj}(\mathfrak{Sts})$
are positive maps $\gamma^{*}:\mathbf{S}'\rightarrow\mathbf{S}$.
$\gamma^{*}$ arises as the dual map of a faithful C*-algebraic
endomorphism $\gamma:\mathcal{A}\rightarrow\mathcal{A}'$ via
\begin{equation}\label{137a}
\gamma^{*}\omega'(A)=\omega'(\gamma(A)),\hspace{0.5cm}\omega'\in\mathbf{S}',\hspace{0.5cm}A\in\mathcal{A}.
\end{equation}
Then a state space for $\mathscr{A}$ is a contravariant functor
$\mathbf{S}$ between $\mathfrak{Man}$ and $\mathfrak{Sts}$:
\begin{equation}\label{E:137b}
\begin{array}{ccc}
(M,\mathbf{g}) & \xrightarrow{\psi} & (M',\mathbf{g}')  \\
\mathbf{S}\downarrow &  & \downarrow\mathbf{S}  \\
\mathbf{S}(M,\mathbf{g}) & \xleftarrow{\alpha^{*}_{\psi}} &
\mathbf{S}(M',\mathbf{g}')
\end{array}
\end{equation}
where $\mathbf{S}(M,\mathbf{g})$ is a set of states on
$\mathscr{A}(M,\mathbf{g})$ and $\alpha_{\psi}^{*}$ is the dual
map of $\alpha_{\psi}$; the covariance property is
$\alpha^{*}_{\tilde{\psi}\circ\psi}=\alpha^{*}_{\psi}\circ\alpha^{*}_{\tilde{\psi}}$
together with the requirement that unit morphisms are mapped to
unit morphisms.

Now let see how can we apply all this formalism to our problem. We
assume there a thermal state on $\mathbb{R}\times\mathbb{R}$. We
also assume it is invariant under the action of the isomorphism,
$\alpha_{t}$, generated by translations in time, the usual time in
Minkowski spacetime.

We would like to proof that when we make $x$, the spatial
coordinate in Minkowski spacetime, 2$\pi$-periodic we still have a
thermal state on the resulting spacetime. Using the structure
given in the diagram (\ref{E:137b}) we just need to proof that
$\alpha_{\psi}$ and $\alpha_{t}$ commute, however as it stand now
we do not know how elements of the algebras in $\mathcal{D}_{c}$
and in $\mathcal{D}_{i}$ are related to each other. Therefore, it
is necessary to introduce more structure before we proof what we
want. Fortunately this structure also has been given by BFV.

We introduce the concept of locally covariant quantum field. This
concept needs the introduction of another category, the category
$\mathfrak{Test}$ of smooth test functions with compact support,
$C^{\infty}_{0}(M)$. The morphisms in this category are the
pushforwards of $\psi$ the morphisms in $\mathfrak{Man}$. Where
$M$ stands for $\mathcal{D}_{c}$ or $\mathcal{D}_{n}$. We also
introduce a family of quantum fields $\Phi_{M,\mathbf{g}}$,
indexed by all spacetimes in $\mathfrak{Man}$. For each spacetime
this field is a map from $C^{\infty}_{0}(M)$ to
$\mathcal{A}(M,\mathbf{g})$
\begin{equation}\label{E:137d}
\Phi_{(M,\mathbf{g})}:C^{\infty}_{0}(M)\rightarrow\mathcal{A}(M,\mathbf{g}).
\end{equation}
This structure can be put in a diagram as
\begin{equation}\label{E:137e}
\begin{array}{ccc}
\mathscr{D}(M,\mathbf{g}) & \xrightarrow{\Phi_{(M,\mathbf{g})}} & \mathscr{A}(M,\mathbf{g})  \\
\psi_{*}\downarrow &  & \downarrow\alpha_{\psi}  \\
\mathscr{D}(M',\mathbf{g}') &
\xrightarrow{\Phi_{(M',\mathbf{g}')}} &
\mathscr{A}(M',\mathbf{g}')
\end{array}
\end{equation}
where the commutativity of the diagram expresses the covariance
for fields
\begin{equation}\label{E:137f}
\alpha_{\psi}\circ\Phi_{(M,\mathbf{g})}=\Phi_{(M',\mathbf{g}')}\circ\psi_{*}.
\end{equation}

Let us now go back to our problem and use the formalism we have
just introduced. Let $f\in\mathscr{D}(M,\mathbf{g})$ and take $M$
as $\mathcal{D}_{c}$ and $M'$ as $\mathcal{D}_{p}$ then, from
(\ref{E:137f}), we have
\begin{equation}\label{E:137g}
\alpha_{\pi^{-1}}\circ\Phi_{(\mathcal{D}_{c},\mathbf{g}_{c})}(f)=\Phi_{(\mathcal{D}_{p},\mathbf{g}_{p})}\circ\pi^{-1}_{*}(f).
\end{equation}
Now, on $\mathcal{D}_{p}$ acts $\Lambda(t)$, the usual translation
in time in Minkowski spacetime. We define a transformation on
$\mathbb{R}^{1}\times\mathbb{S}^{1}$ induced by $\Lambda$ in such
a way that the following diagram commutes
\begin{equation}\label{E:137h}
\begin{array}{ccc}
\mathcal{D}_{c} & \xrightarrow{\pi^{-1}} & \mathcal{D}_{p}  \\
\Lambda'\downarrow &  & \downarrow\Lambda  \\
\mathcal{D}'_{c} & \xrightarrow{\pi^{-1}} & \mathcal{D}'_{p}
\end{array}
\end{equation}
Using $\Lambda$ and $\Lambda'$ we have two maps
\begin{equation}\label{E:137i}
\Lambda_{*}:\mathscr{D}(\mathcal{D}_{p},\mathbf{g}_{p})\rightarrow\mathscr{D}(\mathcal{D}'_{p},\mathbf{g}_{p})
\end{equation}
\begin{equation}\label{E:137j}
\Lambda'_{*}:\mathscr{D}(\mathcal{D}_{c},\mathbf{g}_{c})\rightarrow\mathscr{D}(\mathcal{D}'_{c},\mathbf{g}_{c})
\end{equation}
given by
\begin{equation}\label{E:137k}
\Lambda_{*}f_{p}:=f'_{p}
\end{equation}
and
\begin{equation}\label{E:137l}
\Lambda'_{*}f_{c}:=f'_{c}
\end{equation}
where $f_{p}\in\mathscr{D}(\mathcal{D}_{p},\mathbf{g}_{p})$ and
$f'_{p}\in\mathscr{D}(\mathcal{D}_{p},\mathbf{g}_{p})$, similarly
for $f_{c}$ and $f'_{c}$. The pushforwards induced by $\pi^{-1}$,
$\Lambda$ and $\Lambda'$ are given explicitly by
\begin{equation}\label{E:137m}
\begin{array}{cc}
f_{p}(\mathcal{D}_{p}):=f_{c}(\pi\mathcal{D}_{p}) & f'_{c}(\mathcal{D}'_{c}):=f_{c}(\Lambda'^{-1}\mathcal{D}'_{c})  \\
f'_{p}(\mathcal{D}'_{p}):=f'_{c}(\pi\mathcal{D}'_{p}) &
f'_{p}(\mathcal{D}'_{p}):=f_{p}(\Lambda^{-1}\mathcal{D}'_{p})
\end{array}
\end{equation}
All this structure can be put in the following commuting diagram
\begin{equation}\label{E:137n}
\begin{array}{ccc}
\mathscr{D}(\mathcal{D}_{c},\mathbf{g}_{c}) & \xrightarrow{\pi^{-1}_{*}} & \mathscr{D}(\mathcal{D}_{p},\mathbf{g}_{p})  \\
\Lambda'_{*}\downarrow &  & \downarrow\Lambda_{*}  \\
\mathscr{D}(\mathcal{D}'_{c},\mathbf{g}_{c}) &
\xrightarrow{\pi^{-1}_{*}} &
\mathscr{D}(\mathcal{D}'_{p},\mathbf{g}_{p})
\end{array}
\end{equation}

If also we define the field as a map
$\Phi_{(\mathcal{D}_{c},\mathbf{g}_{c})}:C^{\infty}_{0}(\mathcal{D}_{c})\rightarrow\mathcal{A}(\mathcal{D}_{c},\mathbf{g}_{c})$,
then we have the following commuting diagram
\begin{equation}\label{E:137o}
\begin{array}{ccc}
\mathscr{D}(\mathcal{D}_{c},\mathbf{g}_{c}) & \xrightarrow{\Phi_{(\mathcal{D}_{c},\mathbf{g}_{c})}} & \mathscr{A}(\mathcal{D}_{c},\mathbf{g}_{c})  \\
\Lambda'_{*}\downarrow &  & \downarrow\alpha_{\Lambda'}  \\
\mathscr{D}(\mathcal{D}'_{c},\mathbf{g}_{c}) &
\xrightarrow{\Phi_{(\mathcal{D}'_{c},\mathbf{g}_{c})}} &
\mathscr{A}(\mathcal{D}'_{c},\mathbf{g}_{c})
\end{array}
\end{equation}
Let $f\in\mathscr{D}(\mathcal{D}_{c},\mathbf{g}_{c})$. Then
\begin{equation}\label{E:137p}
\alpha_{\pi^{-1}}\circ\alpha_{\Lambda'}\Phi(f)=\alpha_{\pi^{-1}}\circ\Phi(\Lambda'_{*}f)=\Phi(\pi^{-1}_{*}\circ\Lambda'_{*}f)
\end{equation}
but
\begin{equation}\label{E:137q}
\Phi(\pi^{-1}_{*}\circ\Lambda'_{*}f)=\Phi(\Lambda_{*}\circ\pi^{-1}_{*}f)
\end{equation}
because diagrams (\ref{E:137n}) and (\ref{E:137p}). But
\begin{equation}\label{E:137r}
\Phi(\Lambda_{*}\circ\pi^{-1}_{*}f)=\alpha_{\Lambda}\circ\alpha_{\pi^{-1}}\Phi(f).
\end{equation}
Hence from (\ref{E:137p}), (\ref{E:137q}) and (\ref{E:137r}) we
have
\begin{equation}\label{E:137s}
\alpha_{\pi^{-1}}\circ\alpha_{\Lambda'}\Phi(f)=\alpha_{\Lambda}\circ\alpha_{\pi^{-1}}\Phi(f).
\end{equation}

From diagram (\ref{E:137b}) we see that a positive state on
$\mathscr{A}(\mathcal{D}_{p},\mathbf{g}_{p})$ is mapped to a
positive state on $\mathscr{A}(\mathcal{D}_{c},\mathbf{g}_{c})$.
Also from diagram (\ref{E:137e}), with $M=\mathcal{D}_{c}$,
$M'=\mathcal{D}_{p}$ and $\psi=\pi^{-1}$ we have for
$f\in\mathscr{D}(\mathcal{D}_{c},\mathbf{g}_{c})$
\begin{equation}\label{E:137t}
\Phi(\pi^{-1}_{*}f)=\alpha_{\pi^{-1}}\Phi(f).
\end{equation}
If we denote the state on
$\mathscr{A}(\mathcal{D}_{p},\mathbf{g}_{p})$ as $\omega_{p}$ and
the state on $\mathscr{A}(\mathcal{D}_{c},\mathbf{g}_{c})$ as
$\omega_{c}\equiv\alpha_{\pi^{-1}}^{*}\omega_{p}$ then from
(\ref{137a}) we have
\begin{equation}\label{E:137u}
\omega_{c}(\Phi(f))=\omega_{p}(\Phi(\pi^{-1}_{*}f))=\omega_{p}(\alpha_{\pi^{-1}}\Phi(f)).
\end{equation}

We are assuming that $\omega_{p}$ satisfies the KMS condition,
i.e.,
\begin{equation}\label{E:137v}
\omega_{p}(\Phi(\pi^{-1}_{*}f)\alpha_{t}(\Phi'(\pi^{-1}_{*}g)))=\omega_{p}((\alpha_{t-i\beta}\Phi'(\pi^{-1}_{*}g))\Phi(\pi^{-1}_{*}f))
\end{equation}
where $f$ and $g$ are in
$\mathscr{D}(\mathcal{D}_{c},\mathbf{g}_{c})$. Using
(\ref{E:137s}), (\ref{E:137t}) and (\ref{E:137u}) in
(\ref{E:137v}) we have
\begin{equation}\label{E:137x}
\omega_{c}(\Phi(f)(\alpha_{t}\Phi'(g)))=\omega_{c}((\alpha_{t-i\beta}\Phi'(g))\Phi(f))
\end{equation}
Thus the state $\omega_{c}$ satisfies the KMS condition too.

\section{Disscusion}

It is clear that Haag-Kastler formalism can be applied to a
cylindrical spacetime by replacing Poincar\'{e} symmetry for just
translation symmetry in time and space plus spatial
$2\pi$-periodicity. Then we can consider both quantum field
theories, on the cylinder and on the plane, on the same footing.
The principal differences are the symmetries of the field as
consequences of the manifold symmetries. Now, as we have seen
under the covering map $\pi$ a diamond shaped region in
$\mathbb{R}^{1}\times\mathbb{S}^{1}$ maps to a denumerable
infinite number of diamond shaped regions in
$\mathbb{R}^{1}\times\mathbb{R}^{1}$. Taking into account that an
observable is associated with a local region of spacetime then to
each observable localized in $\mathbb{R}^{1}\times\mathbb{S}^{1}$,
say in $D_{c}$, correspond a denumerable infinite number of
observables localized in $\{D_{i}\}$. The observables in $D_{i}$
are related by an *-isomorphism between the algebras associate to
the family $\{D_{i}\}$. Invoking locality these observables form
an equivalence class given by the equivalence relation $a_{j}\sim
a_{i}$ if $D_{j}=\gamma_{j}D_{i}$ with $\gamma_{j}$ the action of
the abelian translation group. In these circumstances we can
relate a positive state on $\mathbb{R}^{1}\times\mathbb{S}^{1}$ to
a positive state on $\mathbb{R}^{1}\times\mathbb{R}^{1}$ as
follows
\begin{equation}\label{E:140a}
\omega_{BTZ}\left(a\right)\approx\omega_{AdS}\left([a]\right),
\end{equation}
where $[a]$ is the equivalence class associated with the
observable $a$ in $\mathbb{R}^{1}\times\mathbb{S}^{1}$, where
$\approx$ means approximately. We have seen that the formalism
introduced by BFV tell us more precisely how the relation between
states on the cylinder and on the plane should be.

The idea of studying quantum field theory on a multiple connected
spacetime by studying quantum field theory on the covering
spacetime of it is known as automorphic fields \cite{rBasDow79}.
In the previous section we have used the BFV formalism to address
this problem for a simple case, a cylindrical and flat spacetime.
Now we are going to compare it with the image method
\cite{crCrambsKay96} and will show that both formalisms are
equivalent for this case.

Let us consider a scalar quantum field on the flat two dimensional
cylindrical spacetime $\mathbb{R}^{1}\times\mathbb{S}^{1}$. We can
address this problem at least by two procedures. For instance, we
can consider the quantum field on two dimensional Minkowski
spacetime and imposing $2\pi$-periodic boundary conditions. Let us
first consider a $L$-periodic field and later take the particular
case $2\pi$. In this case the two point function turns out to be
\begin{equation}\label{E:140b}
\langle
0_{c}|\hat{\phi}(U,V)\hat{\phi}(U',V')|0_{c}\rangle=-\frac{1}{4\pi}\ln\{(1-e^{-\frac{2\pi}{L}i(U-U'-i\epsilon)})(1-e^{-\frac{2\pi}{L}i(V-V'-i\epsilon)})\},
\end{equation}
where $U=t-x$ and $V=t+x$ are null coordinates and $\epsilon>0$.
Other procedure is to calculate the two point function in
Minkowski spacetime and later use the images sum prescription. The
two point function in Minkowski spacetime is
\begin{equation}\label{E:140b}
\langle
0|\hat{\phi}(U,V)\hat{\phi}(U',V')|0\rangle=-\frac{1}{4\pi}\ln\{(U-U'-i\epsilon)(V-V'-i\epsilon)\}.
\end{equation}
Let us denote the images sum as $F(U,V;U',V')$. Then we have
\begin{equation}\label{E:140c}
F(U,V;U',V')=\sum_{n\in\mathbb{N}_{0}}\langle
0|\hat{\phi}(U,V)\hat{\phi}(U'-Ln,V'+Ln)|0\rangle.
\end{equation}
With the help of the identity \cite{mAbraiaSte65} $\cot
z=\frac{1}{z}+2z\sum_{k=1}^{\infty}\frac{1}{z^{2}-k^{2}\pi^{2}}$,
$z\neq 0, \pm\pi, \pm 2\pi, ...$ we obtain
\begin{equation}\label{E:140d}
F(U,V;U',V')=-\frac{1}{4\pi}\ln\{\sin\frac{\pi}{L}(U-U'-i\epsilon)\sin\frac{\pi}{L}(V-V'-i\epsilon)\}.
\end{equation}
The expression (\ref{E:140b}) can be written as (\ref{E:140d})
plus terms which are linear in $t$ and $t'$ and in $\epsilon$.
Hence the two procedures give the same answer module these terms.
However in a massless two dimensional field theory what really
matters is the two times differentiated two point function
\cite{bsKay00}, hence both procedures give the same answer. These
calculations show the vacuum $|0\rangle$ state is different from
the state $|0_{c}\rangle$. This has been pointed out long time ago
in \cite{bsKay79}. Going back to our problem we can see that by
addressing it with the formalism given by BFV is equivalent to
make the $L$-periodic the field in Minkowski spacetime. Then we
have shown that in this simple case the BFV formalism and
automorphic fields are equivalent.

\vspace{0.5cm}


This work was carried out with the sponsorship of CONACYT Mexico
grant 302006.

\end{document}